\documentclass[]{aa} 

\usepackage{graphicx}
\usepackage{epsfig}
\usepackage{natbib}
\usepackage{savesym}
\usepackage{amsmath}
\savesymbol{iint}
\usepackage{txfonts}
\restoresymbol{TXF}{iint}
\bibpunct{(}{)}{;}{a}{}{,}

\begin{document}

\title{Insights on the Milky Way bulge formation from the correlations
between kinematics and metallicity
  \thanks{Based on ESO-VLT observations 71.B-0617, 73.B-0074, and Paris Observatory GTO 71.B-0196.}
}

\author{C. Babusiaux\inst{1} 
          \and 
          A. G\'{o}mez\inst{1} 
          \and 
          V. Hill\inst{2}
          \and 
          F. Royer\inst{1}
          \and          
          M. Zoccali\inst{3}
          \and 
          F. Arenou\inst{1}
           \and 
	  R. Fux\inst{4}
	   \and 
	  A. Lecureur\inst{1}
           \and 
          M. Schultheis\inst{5}
          \and 
          B. Barbuy\inst{6}     
	  \and          
          D. Minniti\inst{3,8}
	  \and          
          S. Ortolani\inst{7}   
}

\institute{GEPI, Observatoire de Paris, CNRS, Universit\'e Paris Diderot ; 5 Place Jules Janssen 92190 Meudon, France
              \email{Carine.Babusiaux@obspm.fr, Ana.Gomez@obspm.fr, Frederic.Royer@obspm.fr, Frederic.Arenou@obspm.fr}
         \and
              Laboratoire CASSIOPEE, University of Nice Sophia Antipolis, CNRS, Observatoire de la  C\^ote d'Azur, B.P. 4229, 06304 Nice Cedex 4, France
                \email{vanessa.hill@oca.eu}
         \and
             P. Universidad Cat\'olica de Chile, Departamento de Astronom\'\i a y Astrof\'\i sica, Casilla 306, Santiago 22, Chile
                \email{mzoccali@astro.puc.cl, dante@astro.puc.cl}
          \and 
              Observatoire de Gen\`eve, Universit\'e de Gen\`eve, 51 Ch des Maillettes, 1290 Sauverny, Switzerland
          \and
             Observatoire de Besan\c{c}on, CNRS UMR6091, BP 1615, 25010 Besan\c{c}on, France 
                \email{mathias.schultheis@obs-besancon.fr} 
          \and
               Universidade de S\~ao Paulo, IAG, Rua do Mat\~ao 1226,
             S\~ao Paulo 05508-900, Brazil
               \email{barbuy@astro.iag.usp.br}
          \and
             Universita di Padova,Vicolo dell'Osservatorio 5, I-35122 Padova, Italy      
               \email{sergio.ortolani@unipd.it}
	  \and 
	     Vatican Observatory, V00120 Vatican City State, Italy
}

\date{Received ; accepted } 

\abstract
{Two main scenarios for the formation of the Galactic bulge are invoked, the first one through gravitational collapse or hierarchical merging of subclumps, the second through secular evolution of the Galactic disc.} 
{We aim to constrain the formation of the Galactic bulge through studies of the correlation between kinematics and metallicities in Baade's Window ($l=1\degr$, $b=-4\degr$) and two other fields along the bulge minor axis ($l=0\degr$, $b=-6\degr$ and $b=-12\degr$).}
{We combine the radial velocity and the [Fe/H] measurements obtained with FLAMES/GIRAFFE at the VLT with a spectral resolution of R=20000, plus for the Baade's Window field the OGLE-II proper motions, and compare these with published N-body simulations of the Galactic bulge.}
{We confirm the presence of two distinct populations in Baade's Window found in Hill et al. 2010: the metal-rich population presents bar-like kinematics while the metal-poor population shows kinematics corresponding to an old spheroid or a thick disc one. In this context the metallicity gradient along the bulge minor axis observed by Zoccali et al. (2008), visible also in the kinematics, can be related to a varying mix of these two populations as one moves away from the Galactic plane, alleviating the apparent contradiction between the kinematic evidence of a bar and the existence of a metallicity gradient.}
{We show evidences that the two main scenarios for the bulge formation co-exist within the Milky Way bulge.}
 
\keywords{Galaxy: bulge -- Galaxy: formation -- Galaxy: abundances -- Galaxy: kinematics and dynamics}

\maketitle

\section{Introduction}
Although the Milky Way bulge is our closest opportunity to study in detail such a complex chemo-dynamical system, its formation and evolution is still poorly understood. Indeed the high extinction, the crowding, and the superposition of multiple structures along the line of sight make studies of the inner Galactic regions challenging. Two main scenarios have been invoked for bulge formation. The first one is gravitational collapse \citep{Eggen62} or hierarchical merging of subclumps (\cite{Noguchi99}, \cite{Aguerri01}). In this case the bulge formed before the disc and the star-formation time-scale was very short. The resulting stars are old ($>$ 10 Gyr) and have enhancements of $\alpha$ elements relative to iron which are characteristics of classical bulges. The other scenario is secular evolution of the disc through a bar forming a pseudo-bulge (see, e.g. \cite{Combes90}, \cite{Norman96}, \cite{Kormendy04}, \cite{Athanassoula05}). After the bar formation  it heats in the vertical direction (\cite{Combes81}, \cite{Merrifield96}) giving rise to the typical boxy/peanut aspect. 
Observational data of individual stars in our Galaxy provides evidence of both scenarios. The presence of a bar in the inner Galaxy has been first suggested by \cite{Vaucouleurs64} from gas kinematics and confirmed since then by numerous studies including infrared luminosity distribution (e.g. \cite{BlitzSpergel91}), photometric data (e.g. \cite{Stanek94}, \cite{Babusiaux05}, \cite{Benjamin05}), OH/IR and SiO-maser kinematics (e.g. \cite{Habing06}), stellar kinematics (e.g. \cite{Rangwala09a}) and microlensing surveys (e.g. \cite{Alcock00}). Several triaxial structures may even coexist within the Galactic inner regions (\cite{Alard01}, \cite{Nishiyama05}, \cite{LopezC07}). The boxy aspect of the bulge, detected in near-infrared light profile \citep{Dwek95}, also argues for a secular formation of the bulge. 
On the other hand, from medium and high resolution spectroscopic data enhancements of $\alpha$ elements have been observed (\cite{McWilliam94}, \cite{Zoccali06}, \cite{Fulbright07},  \cite{Lecureur07}) suggesting a short formation time-scale.
The discovery of a possible dwarf remnant in the bulge by \cite{Ferraro09} also argues for a hierarchical merging scenario.
A metallicity gradient has been observed in the minor-axis direction of the bulge \citep{PaperI} favouring a bulge formation through dissipational collapse.
Recently some observed similarities between the [$\alpha$/Fe] bulge trend and the trend of the local thick stars (\cite{Melendez08}, \cite{Ryde09}, \cite{AlvesBrito10}) suggest that the bulge and the local thick disc experienced similar chemical evolution histories.
In what concerns the age, colour-magnitude  diagrams show that most of the Galactic bulge stars are older than 10 Gyr (\cite{Ortolani95}, \cite{Feltzing00}, \cite{Zoccali03}, \cite{Clarkson08}) although an intermediate-age population exists (\cite{vanLoon03}, \cite{GroenewegenBlommaert05}).\par

Chemo-dynamical modelling of disc galaxy formation in a Cold Dark Matter (CDM) universe stresses the fact that both types of bulges can coexist in the same galaxy. In the \cite{Samland03}  simulation the bulge contains two stellar populations, an old population formed during the collapse phase and a younger bar population, differing in the [$\alpha$/Fe] ratio. \cite{Nakasato03} suggest that the Galactic bulge may consist of two chemically different components: one rapidly formed through subgalactic clump merger in the proto-Galaxy, and the other one formed later in the inner region of the disc. The [Fe/H] abundance of the merger component tends to be smaller than that of the second component. Recently \cite{Rahimi09}  simulated   bulges formed through multiple mergers and analysed the chemical and dynamical properties of accreted stars with respect to locally formed stars: accreted stars tend to form in early epochs and have lower [Fe/H] and higher [Mg/Fe] ratios, as expected.\par

Exploring correlations of abundances and kinematics in Baade's Window, \cite{Soto07}  suggest  a transition in the kinematics of the bulge, from an isotropic oblate spheroid to a bar, at [Fe/H] $= -0.5$ dex. Their  study was based on 315 K and M giants with proper motions of \cite{Spaenhauer92}, radial velocities of \cite{Sadler96} and low-resolution abundances of \cite{Terndrup95} re-calibrated with the  iron abundances scale of \cite{Fulbright06}. Similar conclusions were obtained  earlier by   \cite{Zhao94} using a sample of 62 K giants 
who pointed out the triaxiality of the bulge from kinematics in  Baade's Window and the fact that the metal poor and metal rich populations were not drawn from the same distribution. \par

In the present paper we analyse the correlations between kinematics and metallicity along  three different minor-axis fields: Baade's Window ($b=-4\degr$), $b=-6\degr$ and $b=-12\degr$,  for which [Fe/H] abundances (\cite{PaperI} hereafter Paper\,I and \cite{PaperII} hereafter Paper\,II) and radial velocities data for about 700 stars have been determined.  Paper\,II obtained [Fe/H] and [Mg/H] metallicity distributions for red clump stars in Baade's Window and showed that the sample seems to be separated into two different populations in the metallicity distributions. We here use the kinematic properties of the samples to confirm and constrain the nature of these two populations in Baade's Window and study their relative proportion change along the bulge minor axis fields of Paper\,I. 
The paper is organized as follows: section 2 summarizes the data used in 
this study. In section 3 we analyse Baade's Window by combining our spectroscopic data with OGLE proper motion data. In section 4 we analyse the radial velocity versus metallicity trend along the bulge minor axis. In section 5 we compare our data with some published N-body models. Section 6 discusses our results.

\section{Data}

The different samples are described in detail in  Paper\,I and in Paper\,II. Only a brief summary is given here.
 Paper\,I samples consist of K giants (called RGB samples in what follows) observed  in the following fields: Baade's Window ($l=1\degr.1$, $b=-4\degr.0$: 194 stars); $b=-6\degr$ ($l=0\degr.2$, $b=-6\degr.0$: 201 stars) and $b=-12\degr$ ($l=0\degr.0$, $b=-12\degr.0$: 99 stars). 
All the stars have photometric $V$, $I$ data coming from different sources (for details see  Paper\,I and Paper\,II) and 2MASS $J$,$H$,$K$ data. 
Spectra  have been obtained with VLT/FLAMES-GIRAFFE spectrograph in Medusa mode  at resolution of about 20\,000. The sample of Paper\,II, observed in the same conditions, contains 219 red clump stars (called RC data in what follows) in Baade's Window ($l=1.0\degr$, $b=-3\degr.9$). Iron abundances have been obtained with a mean accuracy of about 0.2 dex, depending on the value of  [Fe/H], more metal-rich stars have the largest errors. In order to increase the statistics, we have combined the two samples of  Baade's Window (RC sample of Paper\,II and RGB sample of Paper\,I). However due to a difference in the reduction process, the two original MDs were not totally compatible. We therefore used in this combination the new reduction of the RGB sample made with the Paper\,II automatic reduction process (see section 5.1.4 of Paper\,II) which lead to two compatible MDs which we can therefore combine in section \ref{SBW} of this study. However we note that, according to the Besan\c{c}on model (\cite{Robin03}, see figure 9 of Paper\,II and figure 10 of Paper\,I), the two selections have slightly different mean distance and contamination. In section  \ref{Sminoraxis} the other minor axis fields at $b=-6\degr$ and $b=-12\degr$ are compared with the Baade's Window data and, to be consistent, we will use only the original RGB data of Baade's Window computed in Paper\,I in this comparison.\par

Radial velocities have been obtained using the cross-correlation tool available in the GIRAFFE reduction pipeline \citep{Royer02}. The individual spectra were cross-correlated with a box-shaped template corresponding to a K0 giant star. For each target, the barycentric radial velocities derived for each exposure are combined into an average velocity, and the standard deviation is used as an error estimate. For the different fields, the median error on the combined velocities range from 0.26 to 0.43\,km/s. Throughout this paper, $V_r$ stands for the heliocentric radial velocity.

OGLE-II proper motions \citep{Sumi04} are only available for the Baade's Window field. We removed stars with less than 100 data points used in the proper motion computation. Those removed stars are either near the edge of the CCD image or CCD defects or affected by blending. 
In the present paper the mas/yr has been converted in km/s assuming a distance to the Galactic Centre of 8 kpc (1 mas/yr = 37.9 km/s). 
The mean error on the proper motions of our sample is 1.24 mas/yr, i.e. 47 km/s. 
 We worked on the relative values of the proper motions as provided in the catalogue. Section 6 of  \cite{Sumi04} details how one can transform the catalogue into an inertial frame based on the measured proper motions for the Galactic Centre (GC). Our data cover fields BUL\_SC45 and BUL\_SC46, which do not show any difference in the proper motions of the GC (table 3 of  \cite{Sumi04}) so that the relative zero-points from these two fields do not need to be taken into account. \par
 
Throughout this paper errors will be computed from a bootstrap analysis using 1000 samplings of the datasets.
 
\section{\label{SBW}Analysis of Baade's Window}

Our sample of 340 stars in Baade's Window is at present the largest one that has proper motions, radial velocities and homogeneous iron abundances determinations. 
This sample combines both the Baade's Window dataset of Paper\,I (RGB sample) and the dataset of Paper\,II (RC sample) with metallicities derived in an homogeneous way by Paper\,II.
We have computed the proper motion dispersions following the equations in  \cite{Spaenhauer92}:

\begin{equation}
\sigma_\mu^2 = {1 \over (n-1)} \sum_{i=1}^{n} (\mu_i - \bar{\mu})^2 -  {1 \over n} \sum_{i=1}^{n} \xi_i^2
\end{equation}  
with $\mu_i$ and $\xi_i$ being respectively the individual proper motions and their measurement errors. We assume that the measurement errors are uncorrelated. For the radial velocity dispersion computation the errors can be neglected compared to the intrinsic dispersion. 

The covariance $\sigma_{lb}^2$, used in the computation of correlations between the different velocity components and in the the velocity ellipsoid angles, also needs to be corrected from measurement errors due to the transformation from image coordinates to galactic proper motions. Indeed we have:
$$\mu_l = l_{\alpha} \, \mu_{\alpha} + l_{\delta} \, \mu_{\delta}$$
$$\mu_b = b_{\alpha} \, \mu_{\alpha} + b_{\delta} \, \mu_{\delta} $$
with the errors in the equatorial proper motions un-correlated (Sumi 2010, private communication), which implies:
\begin{equation}
\sigma_{lb}^2 = \mathrm{cov}(\mu_l,\mu_b) - {1 \over n} \sum_{i=1}^{n}(l_{\alpha} \, b_{\alpha} \, \xi_{\alpha i}^2 + l_{\delta} \, b_{\delta} \, \xi_{\delta i}^2)
\label{eqslb}
\end{equation}  
with $\xi_{\alpha i}$ and $\xi_{\delta i}$ the individual measurement errors on the equatorial proper motions. In Baade's Window $l_{\alpha} \, b_{\alpha} = - l_{\delta} \, b_{\delta} = -0.426$.

Our Baade's Window radial velocity measurements $\sigma_{r}=111\pm4$ km/s, $\langle V_r \rangle=9\pm6 $ km/s are in excellent agreement with previous studies:  \cite{Howard08}: $\sigma_{r}=112\pm10$ km/s, $\langle V_r \rangle=-5\pm14$ km/s, 
\cite{Rangwala09a}: $\sigma_{r}=112\pm3$ km/s, $\langle V_r \rangle=-1\pm5 $ km/s, 
and other measurements listed in \cite{Rangwala09a}.
The proper motion dispersions $\sigma_{l}=3.13\pm0.16$ mas/yr and $\sigma_{b}=2.50\pm0.10$ mas/yr are also in excellent agreement with the high accuracy $HST$ measurements of \cite{Kozlowski06}: $\sigma_{l}=2.87\pm0.08$, $\sigma_{b}=2.59\pm0.08$ mas/yr.

\subsection{kinematics versus metallicity}

\begin{figure} 
\includegraphics[width=9cm]{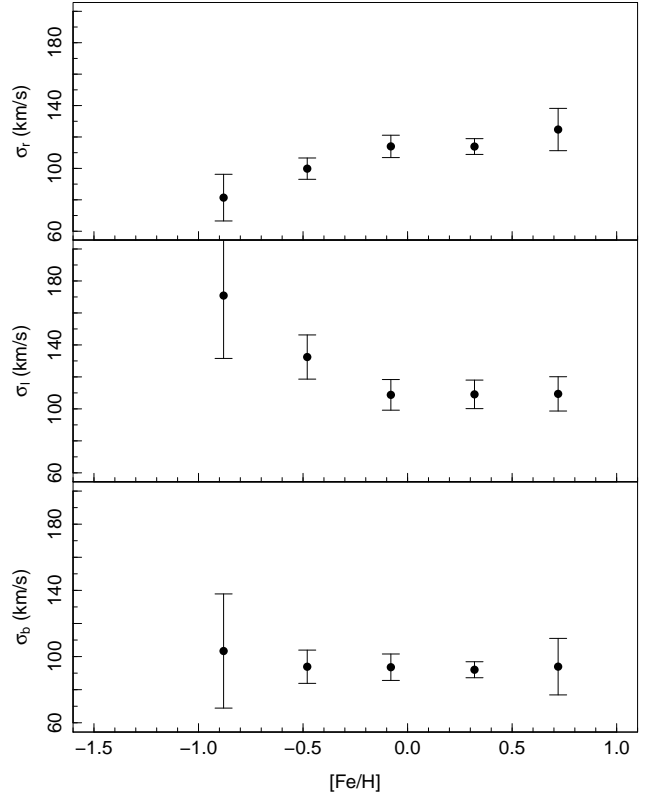}
\caption{Dispersion of the radial ($\sigma_{r}$) and tangential velocities ($\sigma_{l}$, $\sigma_{b}$) as a function of metallicity in Baade's Window by bins of 0.4 dex.}
\label{fig:metdispVelocBW} 
\end{figure}   

Figure \ref{fig:metdispVelocBW} plots the dispersion of velocity components along the line-of-sight ($\sigma_{r}$), the galactic longitude ($\sigma_{l}$) and the galactic latitude ($\sigma_{b}$)  as a function of [Fe/H]. 
The figure shows that $\sigma_{r}$ increases with metallicity, while  $\sigma_{l}$ decreases. No significant variation of $\sigma_{b}$ is seen. 
To quantify if these variations are indeed significant, we use the F-test to compare the variances of the velocity distributions for the first and last 33 \% quantiles ([Fe/H]$<-0.14$ and [Fe/H]$>0.30$): $\sigma_{r}$ and $\sigma_{l}$ are indeed significantly different (p-value = 0.04 and 0.01 respectively), while the variation in $\sigma_{b}$ is not significant (p-value = 0.2). These results do not confirm those of \cite{Soto07}. Their study, based on a compiled sample of  proper motions, radial velocities and low-resolution abundances, showed only a rather shallow variation in $\sigma_{b}$,  in the sense that the more metal poor stars have a higher $\sigma_{b}$ value. The discrepancies with the study of \cite{Soto07} may partly come from the sample itself. Our results are based on more accurate and homogeneous spectroscopic data. The different contamination of the sample by foreground objects may also be another source of discrepancy. Recently,  \cite{Clarkson08} from $HST$ proper motions and estimated photometric parallaxes in the Sagittarius Window constructed the \{$l$,$b$\} velocity ellipse as a function of line-of-sight distance and demonstrated that its properties are sensitive to the distance to the considered objects. They concluded that the depth-integrated velocity ellipsoid of a small population of objects should be treated with caution. Half of our sample are red clump stars while \cite{Soto07} studied M giants, so we expect our sample to be less biased toward closer stars.

\begin{table*}
\caption{Baade's Window velocity ellipsoid parameters. $N$ is the number of stars, $\sigma_{r}$, $\sigma_{l}$, $\sigma_{b}$ are the velocity dispersions, $C_{lr}$, $C_{br}$, $C_{lb}$ are the cross-correlation terms, $l_v=\theta_{lr}$, $\theta_{br}$, $\theta_{lb}$ are the velocity ellipsoid angles. The values are computed for the full data set, the first and the last 33\% quantiles of the metallicity distribution. The values predicted by the dynamical models of Zhao 1996 and Fux 1999, discussed in section 5, are also provided. The velocity ellipsoid parameters of the Fux 1999 model are provided for all the stellar particles as well as separately for the stellar particles originally in the disc and in the old spheroid.}
\label{tab:bwvel}
\centering
\begin{tabular}{crccccccccc} 
\hline\hline
[Fe/H] & $N$ & $\sigma_{r}$ & $\sigma_{l}$ & $\sigma_{b}$ & $C_{lr}$ & $C_{br}$ & $C_{lb}$ & $l_v$ & $\theta_{br}$  &$\theta_{lb}$ \\
(dex) & & (km/s) & (km/s) & (km/s) & &  & & (\degr) & (\degr)  & (\degr) \\
\hline 
all & 340 	& 111 $\pm$ 4 & 119 $\pm$ 6~{ } & { }~95 $\pm$  4 
  & $-0.28 \pm 0.05$  &  0.21  $\pm$  0.06  &  $-0.22  \pm  0.07$	
	& $-38 \pm  5$ & 27  $\pm$  7~{ } & $-22 \pm  6$ \\
$< -0.14$ & 112 & { }~99 $\pm$  6 & 138  $\pm$ 12 & 103 $\pm$ 7 
  & $-0.16 \pm 0.09$  &  0.11  $\pm$  0.10  &  $-0.18 \pm 0.12$
	& $-13  \pm  9$ & 36  $\pm$  19 & $-16 \pm 10$ \\
$>$ { }~0.30 & 111 &  121  $\pm$  6 & 107 $\pm$ 6~{ } & { }~94 $\pm$ 6 
  & $-0.41 \pm 0.09$  &  0.28  $\pm$  0.09  &  $-0.27 \pm 0.10$
	& $-36  \pm  5$ & 24 $\pm$  7~{ } & $-32 \pm 9$ \\
\hline
Zhao96 & - & 113 & 140 & 106 & $-0.21$ & & & & \\
\hline
Fux99 all & 1171 & 122 & 135 & 117 & $-0.13$ &  { }~0.03 & $-0.08$ & $-25$ & { }~10 & $-13$ \\
spheroid & 571 & 122 & 131 & 127 & $-0.01$ & $-0.01$ & $-0.03$ & { }~$-1$ & $-16$ & $-17$ \\
disc         & 600 & 122 & 137 & 107 & $-0.24$ & { }~0.07 & $-0.13$ & $-31$ & { }~13 & $-13$ \\
\hline
\end{tabular}
\end{table*}

In what follows we analyse the shape and the orientation of the velocity ellipsoid for the data sets quoted in Table \ref{tab:bwvel}. From the whole sample, the observed anisotropy in  $\sigma_{l} /  \sigma_{b} = 1.25 \pm 0.09$  is in good agreement with the results of  \cite{Spaenhauer92}, \cite{Feltzing02}, \cite{Kuijken02} and \cite{Kozlowski06}.  We detected also an anisotropy in $ \sigma_{b} /  \sigma_{r} = 0.86 \pm 0.05$ and no significant one in $ \sigma_{l} /  \sigma_{r} = 1.07 \pm 0.07$. Since  the kinematic properties of the whole sample is an average of the properties of  distinct populations, we analyse the shape of the velocity  ellipsoid of the two metallicity samples. 

The metal poor sample ([Fe/H]$<-0.14$) shows a $\sigma_{l}$  higher  than  the other components, $\sigma_{r} \simeq \sigma_{b}$, and an anisotropy in  $\sigma_{l} /  \sigma_{b}\,=\,1.34 \pm 0.17$ (confirmed by a F-test with a 98\% confidence) and in $ \sigma_{l} /  \sigma_{r} =  1.39 \pm 0.16$ (a F-test cannot be performed here due to the large difference in the errors in the proper motion and radial velocity measurements). 
Following \cite{Zhaoetal96}, the observed anisotropies for the metal poor sample ($\sigma_{l} /  \sigma_{b} \simeq  \sigma_{l} /  \sigma_{r} $) may be interpreted in terms of rotation broadening:  as Baade's Window passes close to the minor axis of the bulge, only $\sigma_{l}$ should be broadened by the rotation, which is indeed what is observed. However, a possible contamination with foreground thick disc stars higher than expected would also contribute to an increase in $\sigma_l$.  

For the metal rich sample ([Fe/H]$>0.30$) the shape of the velocity ellipsoid is significantly different from the metal poor sample. We observe for the metal rich population  $\sigma_r > \sigma_l> \sigma_b$, and that $\sigma_{r}$  is higher (with 96\% confidence) and $\sigma_l$ is lower (with 99\% confidence)  than the observed values for the metal poor sample. In the case of a bar, the velocity dispersion along the bar major axis is expected to be much larger than its azimuthal dispersion which is itself lower than its vertical dispersion. If the bar was pointing nearly end-on, that would mean $\sigma_r > \sigma_l> \sigma_b$ which is what we observe. The  anisotropy in $ \sigma_{b} /  \sigma_{r} = 0.78 \pm 0.08$ observed for the metal rich sample may be explained by the flattening of the bar driven population. 

\begin{figure}
\includegraphics[width=10cm]{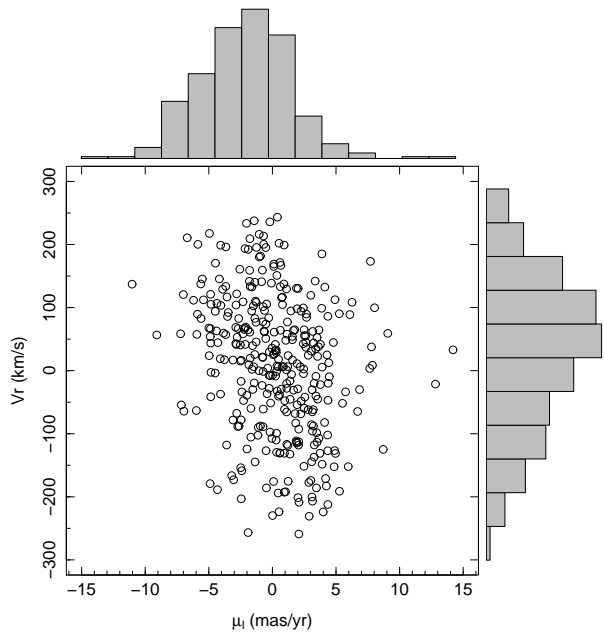}
\caption{$V_r$ and $\mu_l$ distributions in Baade's Window.}
\label{fig:mulVr}
\end{figure}

The Baade's Window velocity ellipsoid shows the characteristics of a non-axisymmetric system. We now analyse the orientation of the velocity ellipsoid. We first examine the correlations between the different velocity components. The obtained cross-correlation terms ($C_{xy} = {\sigma_{xy}^2}/({\sigma_{x}\, \sigma_{y}})$, $\sigma_{xy}^2$ being the covariance) are given in Table \ref{tab:bwvel}. 

 $C_{lb}$ is higher than the results of  \cite{Kozlowski06} ($C_{lb} = -0.10 \pm 0.04$) and \cite{Rattenbury07}  ($C_{lb} = -0.16 \pm 0.03$). We note that \cite{Rattenbury07} used the same OGLE-II data but selected only stars with errors in the proper motion smaller than 1 mas/yr, introducing a bias towards closer stars but reducing the impact of the correction described in equation \ref{eqslb}.

A decrease with metallicity of the correlations between the velocity components seems to be present in our sample.  
We now quantify the observed correlations in terms of velocity ellipsoid angles in Table \ref{tab:bwvel} with:

\begin{equation}
\tan(2 \theta_{xy}) = { 2 \sigma_{xy}^2 \over \vert \sigma_{x}^2 - \sigma_{y}^2 \vert} 
\end{equation}
 
 \begin{figure}
\includegraphics[width=8cm]{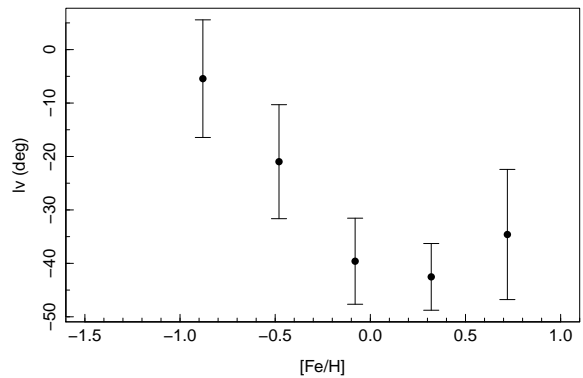}
\caption{Vertex deviation $l_v$ in Baade's Window as a function of metallicity by bins of 0.4 dex.}
\label{fig:metlv}
\end{figure}

In the following we will call the vertex deviation $l_v = \theta_{lr}$, which is the orientation of the axis of the velocity dispersion tensor in the radial-longitudinal velocity plane illustrated in  Fig. \ref{fig:mulVr} for the whole sample. 
Table \ref{tab:bwvel} and Figure \ref{fig:metlv} show that the metal poor sample is consistent with no vertex deviation (Pearson's correlation p-value = 0.1) while the metal rich sample shows a significant vertex deviation ($l_v = -36 \degr \pm 5$ with a Pearson's correlation p-value of $5\times 10^{-5}$).  This confirms the results of \cite{Soto07} and is a clear indication that the metal-rich part of our sample is kinematically under the influence of the bar.  The metal-rich sample may consist essentially of stars formed by the secular disc evolution. Bars are intrinsically non-isotropic, the intrinsic velocity ellipsoid is longer in the radial direction (e.g. \cite{Zhaoetal96}) in agreement with our results. 
Figure \ref{fig:metlv} indicates that the bar influence can already be seen at [Fe/H]$\sim-0.1$ dex. But we note that a small fraction of stars showing a correlation have a large influence on the overall correlation of a sample. 
We used the particles of the \cite{Fux99} model described in section 5 to test that indeed less than 30\% of stars of disc/bar particles are needed to introduce a significant correlation (p-value $<$ 0.05) in the disc/bar plus spheroid sample.

 Our derived value of $\theta_{lb}=-22\degr\pm6\degr$ is consistent with the results of \cite{Clarkson08} ($\theta_{lb}=-34\degr\pm8\degr$) and \cite{Kozlowski06} ($\theta_{lb}=-24\degr$). It shows a significant variation with metallicity. The Pearson's correlation test indicates a correlation with 99\% confidence for the metal rich sample while the metal poor sample correlation is not significant (p-value = 0.1).

We measure for the first time $\theta_{br}$. The Pearson's correlation test does not show a significant correlation between $\mu_b$ and $V_r$ for the metal poor sample (p-value=0.3), while it finds the correlation to be significant  for the metal rich sample with 99.9\% confidence. 

Following Paper\,II, we applied the SEMMUL Gaussian components decomposition algorithm \citep{CeleuxDiebolt86} to the [Fe/H] distribution of the full Baade's Window sample and find a decomposition fully compatible with the Paper\,II results (which were based on the Red Clump sample only): the decomposition identifies clearly two populations, in roughly equal proportions, a metal-poor component (46$\pm$3\% of the sample, $<$[Fe/H]$>$=-0.32$\pm$0.02) with a large dispersion (0.23$\pm$0.01 dex) and a metal-rich component (54$\pm$3\% of the sample, $<$[Fe/H]$>$=0.30$\pm$0.01) with a small dispersion (0.11$\pm$0.01 dex). The variation of the kinematic properties between the metal rich and the metal poor parts of the sample studied here suggests different formation scenarii for those populations. 
The metal poor component of our sample shows  $\sigma_{l}$  higher  than  $\sigma_{r}$  and $\sigma_{b}$, $\sigma_{r} \simeq \sigma_{b}$, a strong anisotropy in $\sigma_l/\sigma_b$ and $\sigma_l/\sigma_r$ and no correlations between the velocity components. This population can be interpreted as a spheroidal population with the velocity anisotropy due mostly to rotation that broadens the observed $\sigma_{l}$ component. 
The metal rich population shows significant anisotropy and correlation between all the velocity components. In particular the strong vertex deviation indicates that it can be interpreted as a bar-driven population.

\subsection{Kinematics versus distance}

We used our red clump stars sample to probe the kinematics against distance. Red clump stars should indeed allow to measure the velocity shift between the two bar streams (\cite{MaoPaczynski02}, \cite{Rangwala09a}). As the stars form, the bar streams in the same sense as the Galactic rotation and the bar being in the first quadrant, the stars on the near side of the bar are expected to go towards us, while stars in the far side should move away from us. The actual velocity shifts between these two streams constrains the bar orientation angle. 

The Besan\c{c}on model confirms that our red clump sample should be nicely centred around the Galactic Centre  (see figure 9 of Paper\,II). 
The magnitude $I$ is less sensitive than magnitude $K$ to both the metallicity and the age for ages older than $\sim$4 Gyr \citep{SalarisGirardi02}. Moreover the $I$ magnitude is more sensitive to the extinction which itself increases with distance, leading the $I$ magnitude to be much more sensitive to distance than $K$. No differential extinction pattern is seen in the region of our sample \citep{Sumi04ext}.

\begin{figure} 
\centering
\includegraphics[width=8cm]{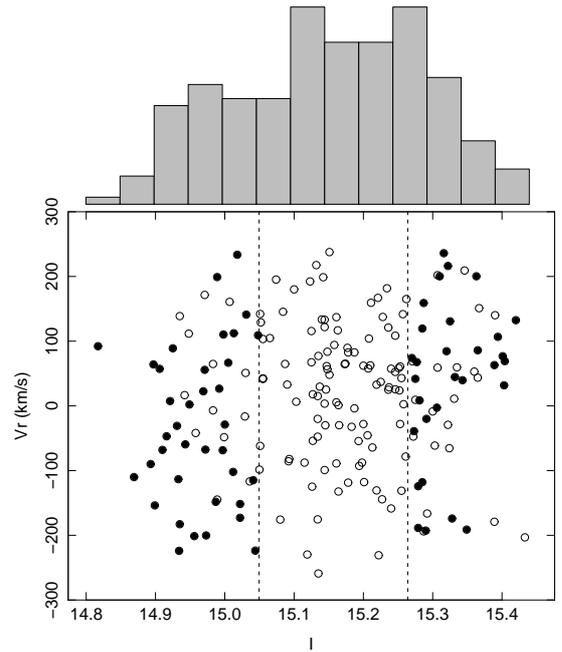}
\caption{Distribution of the $I$ magnitude and $V_r$ in the Baade's Window Red Clump sample. The dotted lines represent the first and last 25\% quantiles of the $I$ magnitude distribution. The filled dots corresponds to stars with [Fe/H]$>-0.09$ used in Table \ref{tab:bwRCvel}}
\label{fig:histI} 
\end{figure}   

\begin{table*}
\caption{Baade's Window Red Clump mean velocities for the faint and bright stars in magnitude $I$ (first and the last 25\% quantiles) for stars with [Fe/H] $>-0.09$ (33\% quantile of the [Fe/H] distribution).  
}
\label{tab:bwRCvel}
\centering
\begin{tabular}{crccc} 
\hline\hline
 & $N$ & ${V_r}$ & ${\mu_l}$ & ${\mu_b}$ \\
 & & (km/s) & (mas/yr) & (mas/yr) \\
\hline 
all & 134  		& { }~{ }~7  $\pm$  10 	&  { }~0.4  $\pm$  0.3  &  $-0.1  \pm$  0.2 \\ 
$I$ $<$ 15.05 & 37 	&  $-32 \pm 20$ 	& { }~0.7  $\pm$  0.5  &  $-0.4 \pm 0.4$  \\
$I$ $>$ 15.26 & 30 	&  { }~38  $\pm$  22 	&  $-0.2  \pm  0.6$  &  { }~0.2  $\pm$  0.5 \\
\hline
\end{tabular}
\end{table*}

We have cut our sample into bright and faint stars on the first and last 25\% quantile of our $I$ magnitude distribution (Fig. \ref{fig:histI}) and computed the mean velocities. Without selection in metallicity, the Welch Two Sample t-test does not indicate that the mean values between the bright and the faint samples are significant enough (p-value = 0.1 for $V_r$, 0.1 for  $\mu_l$ and 0.8 for  $\mu_b$). By removing the metal poor stars (selecting [Fe/H]$>-0.09$, which corresponds to the 66\% quantile of the [Fe/H] distribution) the Welch Two Sample t-test becomes significant on the radial velocities (p-value = 0.02 for $V_r$, 0.2 for  $\mu_l$ and 0.4 for  $\mu_b$). The resulting values are presented in Table \ref{tab:bwRCvel}. The numbers go in the expected direction which is a negative $V_r$ and a positive $\mu_l$ for the near stream and a positive $V_r$ with a negative $\mu_l$ for the far stream. Note that the mean proper motions quoted in Table \ref{tab:bwRCvel} are relative. 
By removing the metal rich stars from the sample (selecting [Fe/H]$<$0.33, which corresponds to the 66\% quantile of the [Fe/H] distribution) the Welch Two Sample t-test does not distinguish differences between the faint and the bright sample (p-value = 0.5 for $V_r$, 0.2 for  $\mu_l$ and 0.9 for  $\mu_b$). 
This again confirms our interpretation of the metal-rich population of Baade's Window being associated to the bar and the metal-poor population to a spheroidal population.  

This shift was not seen in the radial velocities of \cite{Rangwala09a}. They argue that this may be due to Baade's Window population being the superposition of an old spheroidal population and a bar population, which we here confirm. 

For the RGB sample, no correlation of the kinematics with the magnitude $I$ can be seen. We note that the RGB sample is biased towards closest stars. According to the Besan\c{c}on model, the median distance difference between the RC and the RGB sample is expected to be 1 kpc (see figure 9 of Paper\,II for the RC and figure 10 of Paper\,I for the RGB sample). This bias in distance is not significant in the mean kinematics of the RGB sample ($V_r$ = 3 $\pm$  8 km/s, $\mu_l$ = 0.6  $\pm$  0.3 mas/yr,  $\mu_b$ =  0.5 $\pm$  0.2 mas/yr).

\section{ \label{Sminoraxis}Radial velocity versus metallicity along the bulge minor axis}

\begin{figure*}
\begin{minipage}[b]{0.5\linewidth}
\centering
\includegraphics[width=9cm]{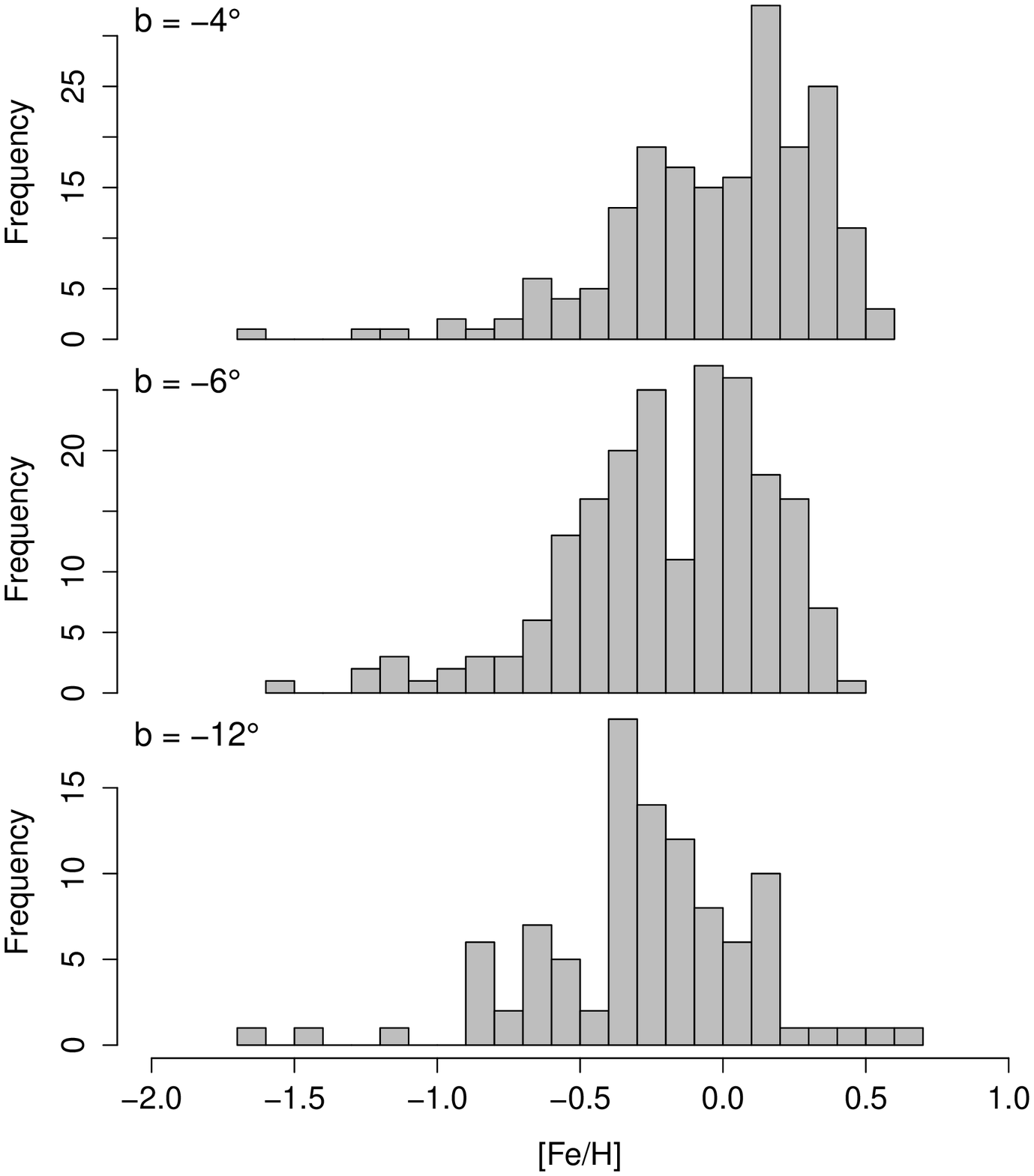}
\caption{Distribution of the metallicity for the different galactic latitudes of Paper\,I}
\label{fig:metdens}
\end{minipage}
\hspace{0.5cm}
\begin{minipage}[b]{0.5\linewidth}
\centering
\includegraphics[width=9cm]{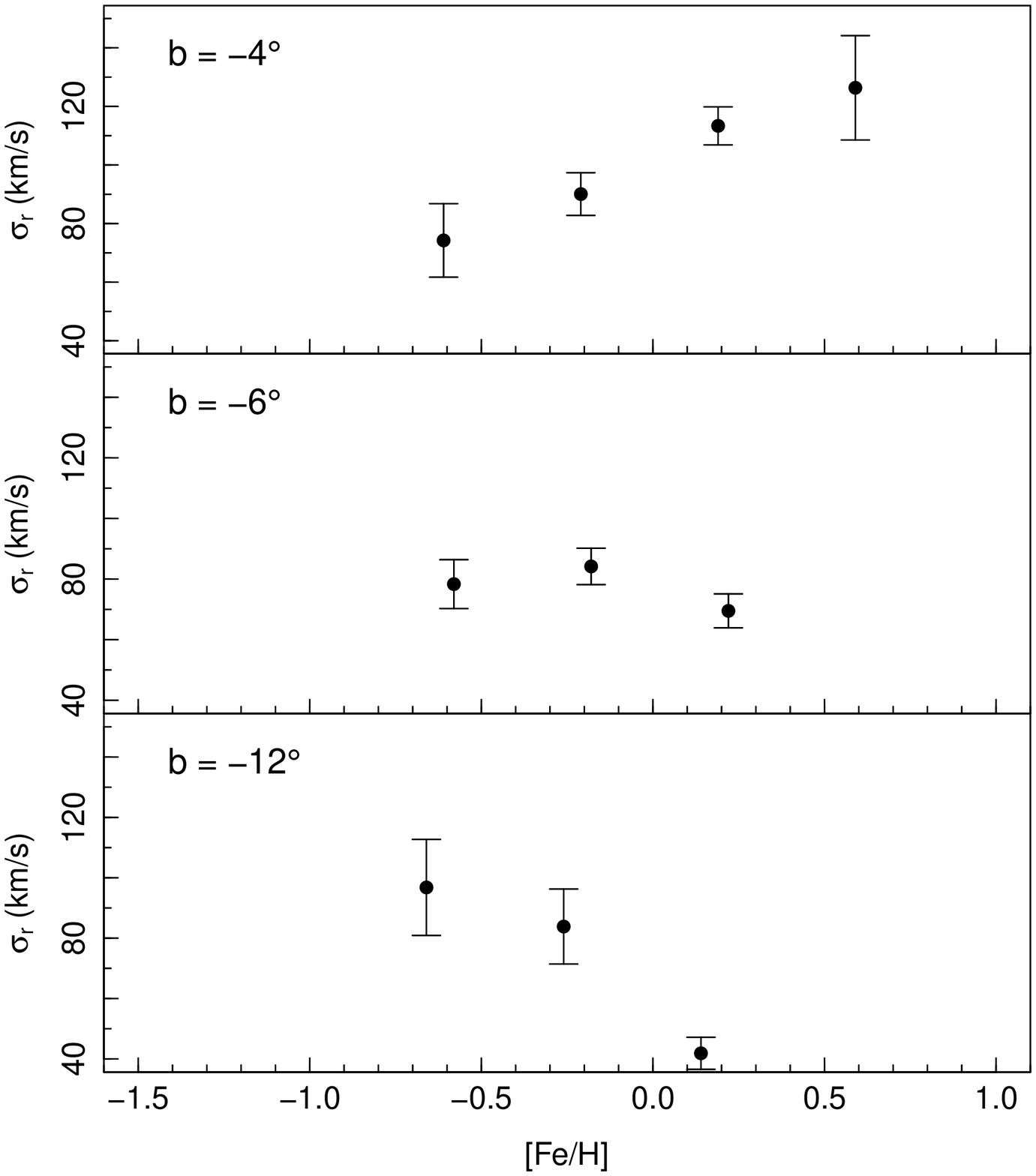}
\caption{Dispersion of the radial velocity for the different galactic latitudes as a function of metallicity by bins of 0.4 dex. }
\label{fig:metdispVr}
\end{minipage}
\end{figure*}

We are now wondering whether the two distinct stellar populations found in the previous section are present all along the
bulge minor axes. We analyse the radial velocity distribution behaviour of the fields of Paper\,I, presented in Table \ref{tab:fieldsSummary}. To  be consistent with the $b=-6\degr$ and $b=-12\degr$ metallicity distribution, we will use only the RGB sample of Baade's Window with the original metallicity distribution function of  Paper\,I.  Paper\,I showed a variation of the metallicity distribution with height $z$ above the Galactic mid-plane (Fig. \ref{fig:metdens}). In this section we combine metallicity data with radial velocity data.  Figure \ref{fig:metdispVr} shows the radial velocity dispersion as a function of metallicity. 

\begin{table*}
\caption{Summary of the bulge minor axis fields (Paper\,I) characteristics.    
$z_{GC}$ is the height of the field along the bulge minor axis, assuming a distance Sun - Galactic Centre of 8 kpc.
$N$ is the number of stars.
$\langle V_r \rangle$ is the mean heliocentric velocity. 
 }
\label{tab:fieldsSummary}
\begin{center}
\begin{tabular}{lrrrrcccc}
\hline\hline
Field & $l$  & $b$ & $z_{GC}$ & $N$ & $\langle V_r \rangle$ & $\sigma_{r}$ & skew & kurtosis\\ 
 & (\degr) & (\degr) & (pc) & & (km/s) &  (km/s) \\
\hline
$b = -4\degr$ & 1.1 & $-4.0$ & 606 & 194 & { }~{ }~5 $\pm$  7 & 104  $\pm$  5 &  $-0.07  \pm  0.1$ &  $-0.54  \pm  0.2$ \\
$b = -6\degr$ & 0.2 & $-6.0$ & 844 & 201 & $-10 \pm 6$ & { }~83 $\pm$ 4 & { }~0.01  $\pm$  0.1 & $-0.19  \pm  0.2$ \\
$b = -12\degr$ & 0.0 & $-12.0$ & 1700 & 99 & $-14 \pm 8$ & { }~80 $\pm$ 8 & { }~0.38  $\pm$  0.4 & { }~1.71  $\pm$  0.7 \\ 
\hline
\end{tabular}
\end{center}
\end{table*}

We first compare the kinematic behaviour along the bulge minor axis (Table \ref{tab:fieldsSummary}) with previous results from the literature. We saw in section 3 that our Baade's Window radial velocity dispersion is in excellent agreement with previous measurements. At $b=-6\degr$ our radial velocity dispersion is lower than the BRAVA measure ($\sigma_{r}=108 \pm 7$ km/s, \cite{Howard08}).
The decrease of the radial velocity dispersion with galactic latitude we observe is consistent with the SiO maser measurements of \cite{Izumiura95paperIII}.

All the fields show a negligible skew, while the kurtosis are different. This variation of the kurtosis was not detected in  \cite{Howard08}. At $b=-12\degr$ the distribution is significantly pointy (with 99.3\% confidence according to the Anscombe-Glynn kurtosis test), indicating that the kinematics are significantly affected by the disc. At $b=-6\degr$ the kurtosis becomes consistent with zero. In Baade's Window the distribution is significantly flattened (with 93\% confidence in the RGB sample and 99.8\% in the red-clump sample). 
\cite{Sharples90} and \cite{Rangwala09a} also measured a skew consistent with zero and kurtosis significantly negative in Baade's Window, indicating that the distributions are flat-topped rather than peaked. \cite{Rangwala09a} concludes that this seems to be consistent with a model of the bar with stars in elongated orbits forming two streams at different mean radial velocities, broadening and flattening the total distribution. 

We analyse now  kinematic data versus metallicity. The mean radial velocities do not show significant variation with metallicity in any of our fields. Figure \ref{fig:metdispVr} shows that the velocity dispersion at the rich end decreases significantly with latitude. However the velocity dispersion on the metal-poor end does not vary significantly with latitude. Summing all stars with [Fe/H] $< -0.5$ dex in all fields leads to $\sigma_{r}=97 \pm 7$ km/s. 

The field $b=-6\degr$ is close enough to Plaut's Window ($l=0\degr$, $b=-8\degr$) so that we can compare our results to the proper motion study of \cite{Vieira07}. The Besan\c{c}on model indicates a mean distance of our selected bulge stars to be 6.5 kpc, compatible with the \cite{Vieira07} selection. Their metallicity distribution is compatible with ours. They do not find neither any change in the proper-motion dispersion as a function of metallicity. They found $\sigma_l=102\pm3$ km/s and $\sigma_b=88\pm3$ km/s at $b=-8\degr$, we obtain $\sigma_r=83$ km/s at $b=-6\degr$, leading to a full picture coherent with the predictions of the model of \cite{Zhao96} of $\sigma_b \sim \sigma_r$ and $\sigma_l > \sigma_b$ at these latitudes (see the bottom of Fig. 6 of \cite{Zhao96}).

We have seen in the previous section that Baade's Window kinematic behaviour as a function of metallicity can be interpreted as a mix of two populations, a metal poor component with kinematics that can be associated to an old spheroid population and a metal rich component with bar-driven kinematics. Under this light we can interpret the variation with galactic latitude of both the metallicity distribution function (Fig. \ref{fig:metdens}) and the kinematic as a function of metallicity (Fig. \ref{fig:metdispVr}) as the bar population disappearing while moving away from the plane. At high latitudes the foreground disc component dominates the metal rich part of the kinematic behaviour. The metal poor component associated with the old spheroid stays present along the bulge minor axis. 

\begin{figure*}
\vspace{0.3cm}
\begin{minipage}[b]{0.49\linewidth}
\centering
\includegraphics[width=8cm]{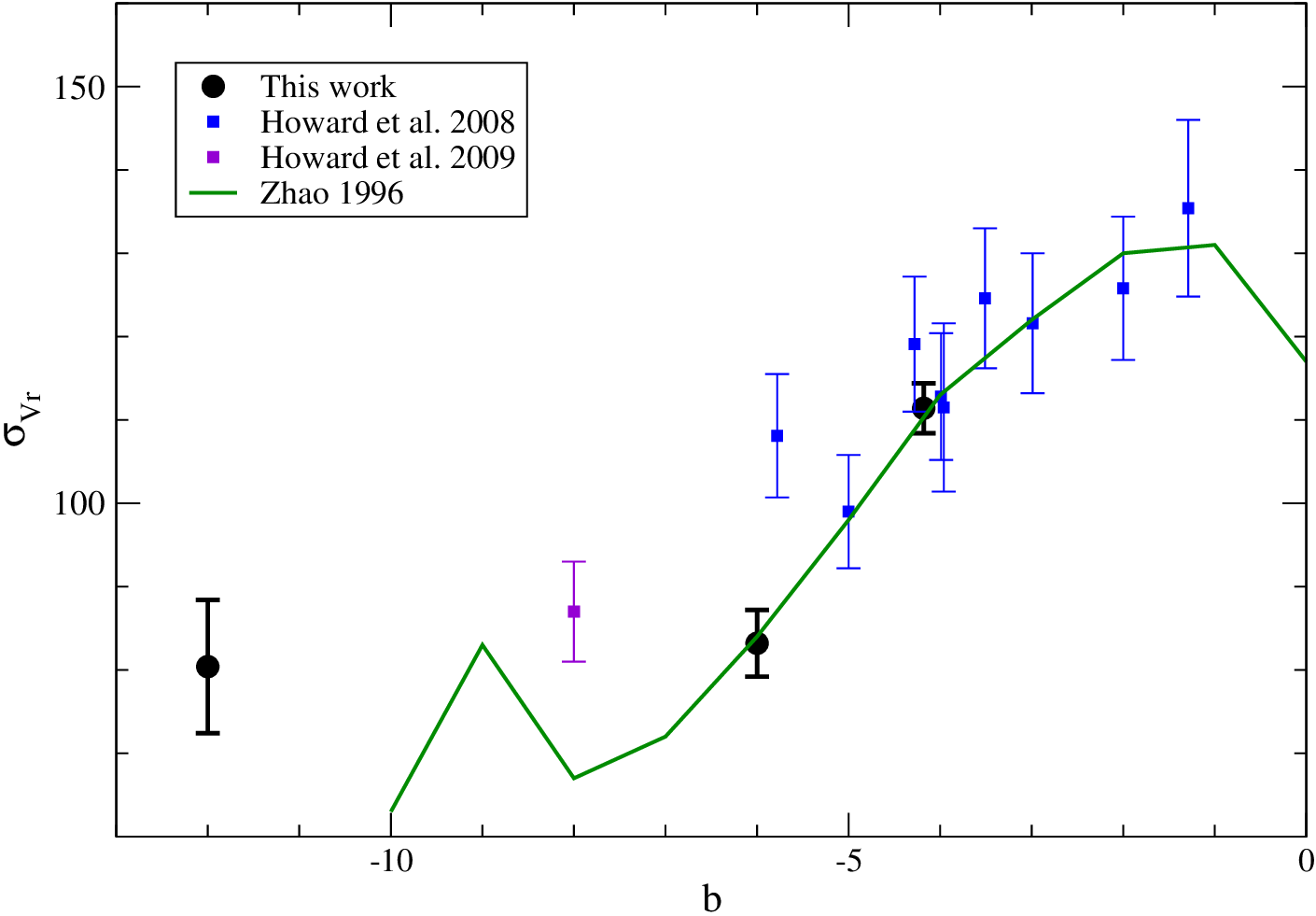}
\caption{Radial velocity dispersion along the bulge minor axis compared to the BRAVA data (\cite{Howard08} and \cite{Howard09}) and the model of \cite{Zhao96}}
\label{fig:compSigVr} 
\end{minipage}
\hspace{0.2cm}
\begin{minipage}[b]{0.49\linewidth}
\centering
\includegraphics[width=8cm]{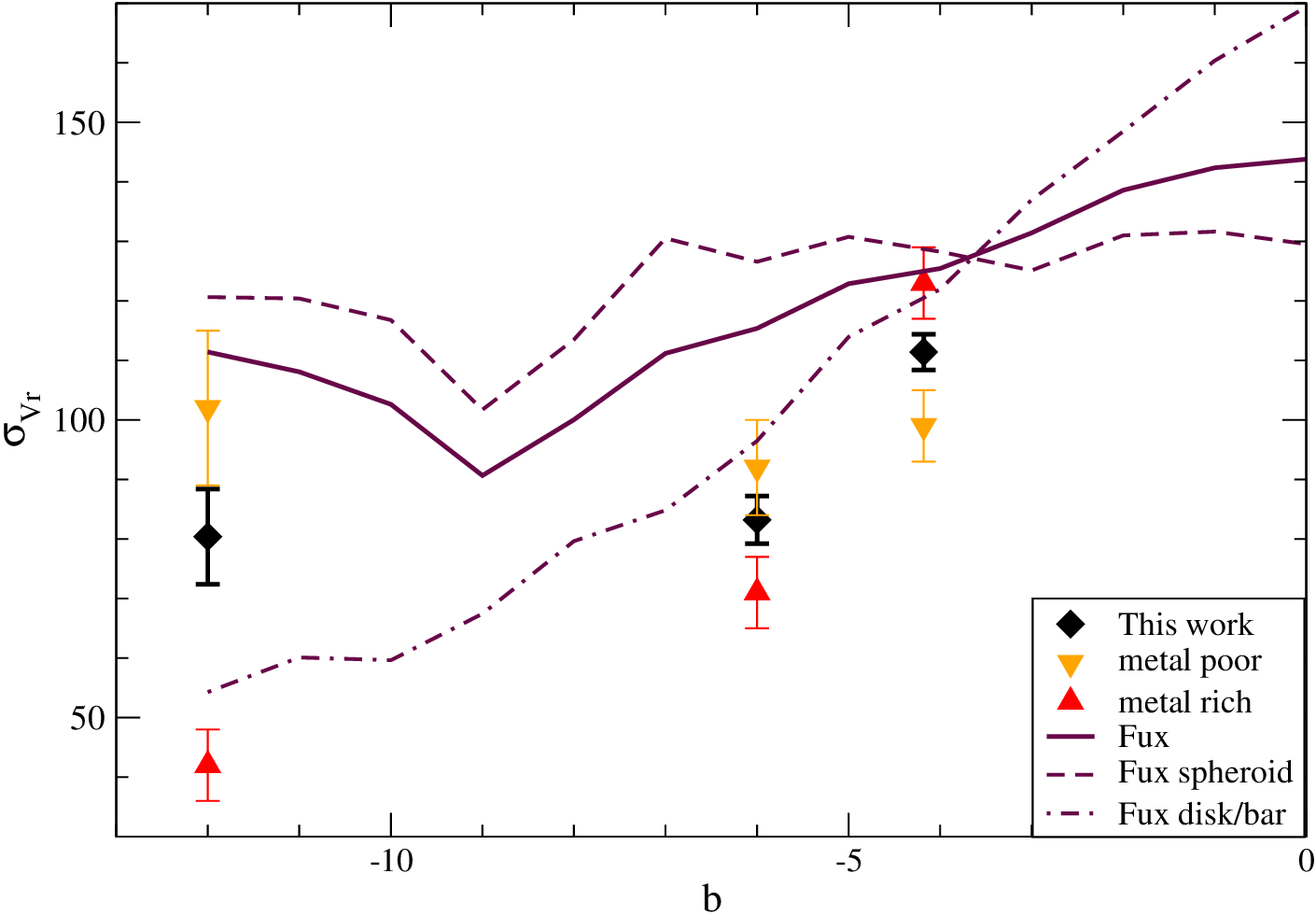}
\caption{Radial velocity dispersion along the bulge minor axis compared with the model of \cite{Fux99}. The metal poor and metal rich population are defined by the 33\% and 66\% quantile of the [Fe/H] distribution.}
\label{fig:compFuxSigVr} 
\end{minipage}
\end{figure*}

We have obtained an estimation of the variation of the populations with latitude by using the SEMMUL Gaussian components decomposition in the metallicity distribution of the Paper\,I samples. 
Table \ref{tab:b4decomp} gives the decomposition of the Baade's Window RGB sample. The proportion of the metal-rich component is higher than in the red clump sample of Paper\,II and its spread in metallicity is higher also, most probably due to a difference in the sample selection which is biased towards stars closer to the Sun in the RGB sample. 
Although both samples are not exactly on the same scale at the high metallicity end (cf section 5.1.4 of Paper II), the mean metallicities for the metal-poor population is extremely similar in both samples (see also the decomposition of the full Baade's Window sample on the same metallicity scale at the end of section 3.1).
At $b=-6\degr$ the Wilks' test allows to keep a solution with 3 components presented in Table \ref{tab:b6decomp} rather than the 2 components one. Population A and population B could correspond to the population A and B observed in Baade's Window. The mean metallicities are coherent although their spread is smaller. The radial velocity dispersions of population A are identical. The radial velocity dispersion of population B decreases at $b=-6\degr$ as expected for a bar-like kinematic behaviour (see Fig. \ref{fig:compFuxSigVr} and associated text). Population C represents only $6\pm2$\% of the sample with a low mean metallicity of $-1.1\pm0.1$ dex and a high velocity dispersion of 127$\pm$26 km/s, which could therefore be associated to the halo. The Besan\c{c}on model prediction of only $0.4\pm0.1$\% of halo star in the sample could therefore have been underestimated. This population could also have been hidden in the population A at $b=-4\degr$. Selecting all stars with [Fe/H]$<-0.9$ in our 3 fields we obtain 16 stars with a radial velocity dispersion of $\sigma_r= 116\pm21$km/s, which is coherent with the solar neighbourhood velocity dispersions measured in this metallicity range (e.g. \cite{ChibaBeers00}: $\sigma_U \sim 110 \pm 10$ km/s) containing both thick disc and halo stars. We are not in a position to clearly associate this population either to the halo or to a metal-poor thick disc in our inner galactic samples.  
SEMMUL did not converge on the $b=-12\degr$ field due to the smaller number of stars and the higher contamination with thin, thick discs and halo stars expected in this field. At $b=-12\degr$ the metal rich component present at $b=-4\degr$ and $b=-6\degr$ seems to have fully disappeared. The metal rich velocity part of the velocity dispersion corresponds to a disc like component (see next section) while the metal poor part shows a velocity dispersion still coherent with the metal poor population of $b=-4\degr$ and $b=-6\degr$, although we cannot distinguish a spheroid and a thick disc contribution.

\begin{table*}
\caption{SEMMUL Gaussian components decomposition of field $b=-4\degr$, RGB sample only. 
The radial velocity dispersion $\sigma_{r}$ is given in km/s.}
\label{tab:b4decomp}
\begin{center}
\begin{tabular}{lllll}
\hline\hline
Pop & [Fe/H] & $\sigma_{[Fe/H]}$ & \% & $\sigma_{r}$ \\ 
\hline
A  & $-0.31 \pm 0.05$ & $0.39 \pm 0.02$ & $33 \pm 3$ & { }~$88 \pm 8$ \\
B  & { }~$0.13 \pm 0.02$ & $0.23 \pm 0.01$ & $67 \pm 3$ & $109 \pm 7$ \\
\hline
\end{tabular}
\end{center}
\end{table*}

\begin{table*}
\caption{SEMMUL Gaussian components decomposition of field $b=-6\degr$. 
The radial velocity dispersion $\sigma_{r}$ is given in km/s. }
\label{tab:b6decomp}
\begin{center}
\begin{tabular}{lllll}
\hline\hline
Pop & [Fe/H] & $\sigma_{[Fe/H]}$ & \% & $\sigma_{r}$ \\ 
\hline
C & $-1.09 \pm 0.07$ & 0.24 $\pm$ 0.01 & { }~6 $\pm$ 2 & 127 $\pm$ 26 \\
A & $-0.27 \pm 0.02$ & 0.24 $\pm$ 0.01 & 64 $\pm$ 3 & { }~83 $\pm$ 5  \\
B &  { }~0.14 $\pm$ 0.02 & 0.13 $\pm$ 0.01 & 30 $\pm$ 3 & { }~70 $\pm$ 6\\ 
\hline
\end{tabular}
\end{center}
\end{table*}

\section{Comparison with dynamical models}

In this section we combine the full Baade's Window sample of section 3 with the b=-6\degr\ and b=-12\degr\ fields of section 4. 

We first compared our kinematics with the model of \cite{Zhao96} in Baade's Window  (Table \ref{tab:bwvel}) and along the minor axis (Fig. \ref{fig:compSigVr}). The model of \cite{Zhao96} is a 3D steady-state stellar model using a generalized Schwarzschild technique consisting of orbital building blocks within a bar and a disc potential. We find a very good agreement between our data and this model. Figure \ref{fig:compSigVr} shows also the very good agreement of our radial velocity distribution along the minor axis with the BRAVA data (\cite{Howard08}, \cite{Howard09}).
These comparisons are done on global kinematics while we have shown here that several populations are present in these data.    

The N-body dynamical model of \cite{Fux99} allows to compare the kinematic properties of the particles that were originally in the disc to those being in the old spheroid. This model is a 3D self-consistent, symmetry-free N-body and smooth particle hydrodynamics code. It contains 3.8 million particles: a dark halo, a disc and a spheroid and a gas component. 
Following \cite{Howard09} we have used the Fux model c10t2066 which assumes a bar angle of 20\degr\ and a distance to the Sun of 8 kpc. We have selected particles at a distance from the Sun $6<r<10$ kpc in a cone selection of 0.5 degree radius for $\vert b\vert<7\degr$ and in a cone selection of 1 degree for $\vert b\vert>7\degr$ to allow to work with a sample large enough away from the plane (this larger cone leads to 209 particles at $b=-12\degr$). We computed statistical uncertainties of the model by bootstrap of 2-3 km/s in the velocity dispersions and 5$\degr$ in the velocity ellipsoid angles for the disc/bar, 20$\degr$ for the spheroid. 

Table \ref{tab:bwvel} shows the predictions of the model of \cite{Fux99} in Baade's Window. In this field the model contains 50\% of stars from the spheroid and 50\% from the disc/bar population, equivalent to the SEMMUL Gaussian components decomposition we obtained for the red-clump sample in Paper\,II. The velocity dispersions of the model are slightly higher than our observed ones and the velocity components correlations are lower.  The velocity dispersions of the metal rich population corresponds to the disc/bar particles except for $\sigma_l$ which is higher in the model. Both the metal rich observations and the disc/bar particles show a correlation of the velocity components but the correlations in the observations are higher. The vertex deviation is however identical in the model and in the observations. The metal poor component show lower velocity dispersions in $\sigma_r$ and $\sigma_b$ than the spheroid model particles. As expected the model does not show any correlations in the velocity components of the spheroid particles, in line with our metal poor sample. 

Figure \ref{fig:compFuxSigVr} shows the strong decrease in radial velocity dispersion predicted by the Fux model for the disc/bar population while moving away from the plane. In the plane the predicted dispersion is larger for the disc/bar than the spheroid one, due to the influence of the bar driven orbits. At high latitude the $\sigma_r$ dispersion is small for the bar/disc component, following a disc like behaviour. The spheroid population keeps the same velocity dispersion along the minor axis. 
In Fig. \ref{fig:compFuxSigVr} we have over-plotted the Fux model radial velocity dispersions with our data for the full sample, the metal rich and the metal poor parts, as defined by the 33\% and 66\% quantiles in the [Fe/H] distribution. For the metal rich part we observe a strong decrease of the metallicity dispersion while going away from the plane, coherent with the behaviour of the disc/bar particles of Fux model. For the metal poor part we observe a constant velocity dispersion, as for the spheroid particles of the Fux model, but with a mean velocity dispersion of about 20 km/s lower than the model. It is this difference of velocity dispersion for the metal poor component which leads to a too high global velocity dispersion of the Fux model compared to the BRAVA data, leading \cite{Howard09} to an interpretation different from ours: at $b=-8\degr$ the radial velocity distribution is compatible with the disc/bar component of the Fux model without the need for an old spheroid component. However with a velocity dispersion for the spheroid of $\sim$100 km/s instead of $\sim$120 km/s the global velocity dispersion would be coherent with the full Fux model (spheroid+disc). 
The use of the Besan\c{c}on model adds support to the interpretation that the decrease in $\sigma_{r}$ with latitude, seen in disc/bar particles of \cite{Fux99} and in our metal-rich samples, can be due to the disc replacing the bar in the sample: according to the Besan\c{c}on model, the thin disc contamination of our sample at $b=-6\degr$ was expected to be of 10\% with a $\sigma_{r}$ = 62 km/s and 19\% at $b=-12\degr$ with $\sigma_{r}$ = 47 km/s. 

\section{Discussion}

Our analysis of the kinematics as a function of metallicity in Baade's Window shows that our sample can be decomposed in two distinct populations for which we suggest different formation scenarii.
The metal poor component does not show any correlation between its velocity components and is therefore consistent with an isotropic rotating population. Paper\,II showed that this component is enriched in [Mg/Fe]. We interpret this population as an old spheroid with a rapid time-scale formation.
The metal rich component shows a vertex deviation consistent with that expected from tags a population with orbits supporting a bar. Paper\,II showed that this component has a [Mg/Fe] near solar.
We interpret this population as a pseudo-bulge formed over a long time scale through disc secular evolution under the action of a bar. 
This pseudo-bulge is gradually disappearing when moving away from the Galactic plane.

In this context we can give a new consistent interpretation of the metallicity gradient in the bulge. A metallicity gradient is indeed visible in the bulge when observing further away from the plane than Baade's Window (\cite{Frogel99} and Paper\,I), while in the inner regions ($\vert b  \vert\leq4\degr$) no gradient in metallicity has been found (\cite{Ramirez00} and \cite{Rich07}). This can be understood if the bar as well as the old spheroid population are both present in the inner regions, leading to a constant metallicity at $\vert b \vert\leq4\degr$, while the bar influence gradually fades further away from the plane than Baade's Window. At high latitudes only the old spheroid remains: the mean metallicity of the outer bulge measured by \cite{IbataGilmore95} of [Fe/H]$\sim-0.3$ dex corresponds very well to our metal poor population. This scenario is also consistent with the distribution of bulge globular clusters of \cite{Valenti09} who found no evidence for a metallicity gradient but all their clusters with [Fe/H]$>-0.5$ dex are located within ($\vert b  \vert\leq5\degr$). 

In what concerns the age of the two populations, we expect the spheroid component to be old while the pseudo-bulge component may contain both the old stars of the inner disc redistributed by the bar and younger stars whose formation has been triggered by the bar gas flow. 
\cite{vanLoon03} found that although the bulk of the bulge population is old, a fraction of the stars are of intermediate age (1 to 7 Gyr).  
\cite{GroenewegenBlommaert05} observed Mira stars of ages 1-3 Gyr at all latitudes from $-1.2$ to $-5.8$ in the OGLE-II data.
\cite{Uttenthaler07} found 4 bulge stars with ages lower than 3 Gyr at a latitude of $b=-10\degr$. 
\cite{Bensby09} found 3 microlensed bulge dwarfs with ages lower than 5 Gyr. 
87\% of the variable stars detected by \cite{KouzumaYamaoka09} are distributed within $\vert b  \vert\leq5\degr$ and most of them should be large-amplitude and long-period variables such as Mira variables or OH/IR stars. This intermediate age population has been shown to trace the Galactic bar (\cite{vanLoon03}, \cite{Izumiura95paperIII}, \cite{GroenewegenBlommaert05}, \cite{KouzumaYamaoka09}), although providing a larger bar angle ($\sim40\degr$) than studies based on older tracers such as red clump stars ($\sim20\degr$, \cite{Stanek94}, \cite{Babusiaux05}). This discrepancy in the bar angle could well be explained if the old tracers probed a mix of spheroid and bar structures while the young tracers only probe the bar one (although biases on the longitude area surveyed needs also to be taken into account, see \cite{Nishiyama05}). 
If this intermediate age population was associated to a part of the bar component, their presence in the CMDs would decrease while going away from the plane as the main bar component and would therefore be a small fraction of the CMD of \cite{Zoccali03} at $b=-6\degr$. 
\cite{Clarkson08} obtained a proper motion decontaminated CMD with a well defined old turn-off in an inner field ($l=1\degr$, $b=-3\degr$). However we would expect an intermediate age population associated with the bar to be metal rich, which, due to the age-metallicity degeneracy, would imply that this population could be hidden in the CMD of \cite{Clarkson08} if its contribution is small enough compared to the bulk of the bulge population. The new filter combination proposed by the ACS Bulge Treasury Programme to break the age-metallicity-temperature degeneracy \citep{Brown09} should provide new lights on this issue.

We note that in Baade's Window, neither the kinematics nor the chemistry allows to distinguish what we call the old spheroid to the thick disc. The mean metallicity of the solar neighbourhood thick disc is however lower (e.g. \cite{Fuhrmann08} derived [Fe/H] $=-0.6$ and [Mg/H] $=-0.2$) than the mean metallicity of our metal poor population ([Fe/H]$= -0.27$ dex and [Mg/H]$=-0.04$, Paper\,II). 
\cite{Melendez08}, \cite{Ryde09}, \cite{Bensby09} and \cite{AlvesBrito10} observed similarities between the metallicity of the bulge and the metallicity of thick disc stars for metal poor stars. The sample of \cite{Ryde09} contains 11 of our stars all with less than solar metallicity. 
Simulations of the formation of thick stellar discs by rapid internal evolution in unstable, gas-rich, clumpy discs \citep{Bournaud09} show that thick discs and classical bulges form together in a time-scale shorter than 1 Gyr which allow to explain the observed abundance similarities.

The coexistence of classical and pseudo bulge has been observed in external galaxies (\cite{Prugniel01}, \cite{Peletier07}, \cite{Erwin08}) and obtained by N-body simulations (\cite{Samland03}, \cite{Athanassoula05}). 
The chemical and dynamical model of \cite{Nakasato03} suggests the presence in the bulge of two chemically different components as we found, one formed quickly through the subgalactic clump merger in the proto-Galaxy, and the other has formed gradually in the inner disc. But they do not have a bar in their model. They fitted well the kinematics of \cite{Minniti96} who observed at $l=8\degr$, $b=7\degr$ a decrease of $\sigma_r$ with metallicity, which is coherent with what we observed at $b=-12\degr$ and the fact that \cite{Nakasato03} defined the bulge radius as R$<$2kpc. 
The chemo-dynamical model of \cite{Samland03} predicts the different characteristics of our sample: their total bulge population contains two stellar populations: a metal rich population ([Fe/H]$>$0.17) with [$\alpha$/Fe]$<$0 associated with the bar, and an old population that formed during the proto-galactic collapse, with a high [$\alpha$/Fe] and a [Fe/H] corresponding to the ``thick disc'' component. Their model also predicts the resulting apparent metallicity gradient along the bulge minor axis (their Fig. 13). 

Our study highlights the importance to combine metallicity to 3D-kinematic information to disentangle the different bulge populations. This approach needs to be extended to various galactic longitudes. Gaia will not only provide those but also allow to determine the distances (probing the different structures along the line of sight and removing distance induced biases) and work on an impressively large sample of un-contaminated bulge stars. 

\begin{acknowledgements}
MZ and DM are supported by FONDAP Center for Astrophysics 15010003, the BASAL Center for Astrophysics and Associated Technologies PFB-06, the FONDECYT 1085278 and 1090213, and the MIDEPLAN MilkyWay Millennium Nucleus P07-021-F.
\end{acknowledgements}

\bibliographystyle{aa} 
\bibliography{BulgeKine}

\end{document}